\documentclass[pss,fleqn,embeddedheads]{w-art}
\usepackage{cite}
\usepackage{times}
\usepackage{w-thm}
\usepackage[]{graphicx}

\begin{document}


\renewcommand{\copyrightyear}{2006}
\DOIsuffix{theDOIsuffix}
\Volume{XX}
\Issue{1}
\Month{06}
\Year{2006}
\pagespan{1}{}
\Receiveddate{\sf zzz} \Reviseddate{\sf zzz} \Accepteddate{\sf
zzz} \Dateposted{\sf zzz}


\title[Coherent destruction of the current]{Coherent destruction of the current through molecular wires using short laser pulses}


\author[uk]{Ulrich Kleinekath\"ofer \footnote{Corresponding
     author: e-mail: {\sf kleinekathoefer@physik.tu-chemnitz.de}, Phone: +49\,371\,531\,33147, Fax:
     +49\,371\,531\,3151}\inst{1,2}}
\address[\inst{1}]{Institut f\"ur Physik, Technische Universit\"at Chemnitz, 09107
  Chemnitz, Germany}
\address[\inst{2}]{International University Bremen, P.O.Box 750 561, 28725
  Bremen, Germany}
\author[gq]{GuangQi Li\inst{1}}
\author[sw]{Sven Welack\inst{1,3}}
\author[ms]{Michael  Schreiber\inst{1}}
\address[\inst{3}]{Department of Chemistry, Hong Kong University of Science
  and Technology, Kowloon, Hong Kong} 
\begin{abstract}
  A molecular wire coupled to two electron reservoirs is investigated
  within a tight-binding approach including spin and Coulomb interaction.
  Under the assumption of weak coupling to the electron reservoirs a
  quantum master equation can be derived for the electron transport through
  the wire.  Motivated by the phenomenon of coherent destruction of
  tunneling for monochromatic laser fields, the influence of Gaussian laser
  pulses on the transport through the wires is studied.  For situations in
  which the maximum amplitude of the electric field fulfills the conditions
  for the destructive quantum effect the average current through the system
  can be suppressed even for a wire consisting of only one site. Turning on
  the electron correlation does not destroy the suppression of the current
  by the laser.
\end{abstract}
\maketitle                   




\renewcommand{\leftmark}
{U. Kleinekath\"ofer et al.: Coherent destruction of the current through molecular wires using short laser pulses}

\section{Introduction}

Electronic transport through molecular wire has been studied intensively
theoretically as well as numerically in recent years
\cite{hang02b,nitz03a}.  A molecular wires consists of a molecule which is
coupled to two leads acting as electron source and drain.  Depending on the
bias voltage applied across the wire there will be a charge flow through
the wire which might consist of a single molecule only.  Most electron
transport experiments are described by the scattering approach put forward
by Landauer \cite{land57,datt95}.  Another class of theories of quantum
transport utilizes the non-equilibrium Green's function approach
\cite{meir92}. This formalism has the advantage of being formally exact
within the lead-wire coupling but the dependence of Green's functions on
two time arguments makes it rather difficult to calculate the quantities
involved. If one treats the lead-wire coupling perturbatively one can
derive quantum master equations for the electron dynamics in the wire and
the current through the wire which can be treated in the standard fashion
for master equations \cite{kohl04a,li05,ovch05a,wela05a}.  This kind of
quantum master equations is usually used to study the dynamics of a system
coupled to a thermal bath \cite{blum96,may00}.  Often these theories are
based on a second-order perturbation theory in the system-bath coupling and
sometime together with the neglect of memory effects (Markov
approximation). In the present contribution we will use this formalism for
the coupling of the wire to fermionic reservoirs \cite{wela05a,wela05b}.
We make use of the spectral decomposition of the spectral density
$J(\omega)$ first introduced by Meier and Tannor \cite{meie99}.  Initially
this formalism was applied within the time-nonlocal approach based on the
Nakajima-Zwanzig equation.  Afterwards the same methods was used for master
equations based on a time-local formalism \cite{yan05,klei04a}. For
molecular wires these two versions of quantum master equations have been
compared recently\cite{wela05b}.

In addition to the coupling to  fermionic reservoirs,
the relevant system, i.e.\ the wire, is coupled to a laser field.  First
experimental \cite{gers00} and theoretical \cite{kohl04a,wela05a} studies
in this direction have been performed. On the theoretical side most of
these studies are based on a perturbative treatment in the environmental
coupling this allows for an easy inclusion of time-dependent laser
fields.  For time-periodic laser fields the Floquet formalism can be
employed \cite{kohl04a} while for arbitrary time-dependent laser fields the
present authors derived a formalism which treats the laser-wire interaction
exactly \cite{wela05a}.  This approach employs the above described spectral
decomposition of the spectral density.

In previous publications \cite{lehm03a,lehm04a,wela05a} it was shown that
the current through the molecular wire can be suppressed using the
phenomenon of coherent destruction of tunneling (CDT). This effect was
first studied by Grossmann et al.\cite{gros91,gros92} showing that the
tunneling dynamics in a periodically driven quantum system can be quenched.
At a fixed frequency of the laser the CDT occurs for certain amplitudes of
the laser field. In the case of periodic laser fields the CDT can be
studied using the Floquet theory \cite{lehm03a,lehm04a}. In a recent study
\cite{klei06b} we showed that the CDT can also be observed for rather short
laser pulses.  In the case of a cw-laser field one has to average the
current over one period of the carrier frequency to see the complete
suppression of the current. For short laser pulses the current has to be
averaged over several periods. In Ref.~ \cite{klei06b} three periods were
employed for this purpose, here we take five periods.  In the current
contribution we want to extend the previous study by including spin to be
able to study on-site electron correlation. Furthermore we study the
effects of the laser pulse on a system with one site only. Throughout the
paper $\hbar$ and the Boltzmann constant $k_B$ are set to unity.

\section{Hamiltonian and master equation}

\begin{figure}
\centerline{\includegraphics[width=5.5cm,clip]{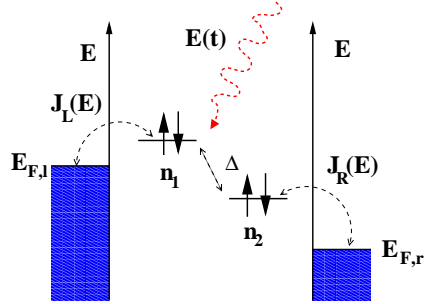}}
\caption{The two-site system  is coupled
  to two electronic reservoirs with Fermi energies $E_{F,l}$ and $E_{F,r}$
  and at the same time to a thermal bath.  The coupling of the sites to
  each other is described by the hopping element $\Delta$ and to the
  corresponding lead by the spectral density function $J_R(\omega)$. The
  on-site energies $E_n$ of the wire can be manipulated with a
  time-dependent external electric field $E(t)$.}
\label{f.0}
\end{figure}

To study the effects mentioned in the introduction we use 
a system consisting of two parts:  the relevant system $H_S(t)$,
mimicking the wire, and the  reservoirs $H_R$ modeling the leads.
Thus the total Hamiltonian is given by  
$ \label{equ:Ham_total}
H(t)=H_S(t)+H_R+H_{SR} $
with the wire-lead coupling $H_{SR}$.  The
wire consists of $N$ sites coupled to each other by a hopping element
$\Delta$ (see Fig.~\ref{f.0}).  The creation and annihilation operators of
electrons at site $n$ with spin $\sigma$ are denoted by
$c^\dagger_{n\sigma}$ and $c_{n\sigma}$, respectively, so that the
tight-binding description of the the molecular wire reads
\begin{equation} \label{equ:Ham_wire}
H_S(t)=\sum_{n\sigma} E_n c_{n\sigma}^\dagger c_{n\sigma} - \Delta
\sum_{n\sigma} ( c_{n\sigma}^\dagger c_{n+1,\sigma} 
+c_{n+1,\sigma}^\dagger c_{n\sigma})
+ U \sum_{n}   c_{n\uparrow}^\dagger c_{n\uparrow}  
c_{n\downarrow}^\dagger c_{n\downarrow}  - \mu{}E(t)~. 
\end{equation}
In the first term we assumed the on-site energies $E_n$ to be
spin-independent.  The second term describes the nearest-neighbor hopping
of electrons with the same spin and the third term the on-site electron
interaction within the wire. The laser-wire interaction is given by the
last term and for the dipole operator we assume \cite{kohl04a}
\begin{eqnarray}
  \label{eq:1}
  \mu{}= e\sum_n x_n  = e \sum_{n\sigma} \frac{2n - N -1}{2} 
 c_{n\sigma}^\dagger c_{n\sigma}~
\end{eqnarray}
for wires with more than one site and a constant for a wire consisting of
only one site.

The two electronic leads that are modeled by two independent reservoirs of
uncorrelated electrons in thermal equilibrium. In the derivations below
only the left lead will be considered but the formalism has to be applied
to the right lead as well.  Denoting the creation and annihilation
operators of an electron in the reservoir modes $\omega_q$  with spin
$\sigma$ by $c_{q\sigma}^\dagger$ and $c_{q\sigma}$, respectively, the
coupling of the left lead to first site of the wire is given by
\begin{equation} \label{equ:Ham_coup}
H_{SR}= \sum_{\sigma,x=1,2} K_{x\sigma}  \Phi_{x\sigma} 
=  \sum_{q\sigma} (V_q c_{1\sigma}^\dagger c_{q\sigma} 
+ V_q^* c_{q\sigma}^\dagger c_{1\sigma})
\end{equation}
with $\Phi_{1\sigma}=\Phi_{2\sigma}^{\dagger}=\sum_q V_q c_{q\sigma}$,
$K_{1\sigma}=K_{2\sigma}^{\dagger}=c_{1\sigma}^\dagger$, and a wire-lead
coupling strength $V_q$

As one is normally not interested in the dynamics within the leads but only
within the wire, a quantum master equation (QME) based on a second-order
perturbation theory in the wire-lead coupling has been developed
for the reduced density matrix of the wire $\rho_S(t)$ \cite{wela05a} 
\begin{eqnarray}\label{equ:master2local}
\frac{\partial\rho_S(t)}{\partial t}&=&-i \mathcal L_S(t) \rho_S(t) 
 -\sum_{\sigma xx'} [K_{x\sigma},\, \Lambda_{xx'\sigma}(t)\rho_S(t)-
\rho_S(t)\widehat \Lambda_{xx'\sigma}(t)]
\end{eqnarray}
with the corresponding auxiliary operators for the wire-lead coupling
\begin{equation}\label{equ:aux1local}
\Lambda_{xx'\sigma}^{}(t) = \int_{t_0}^t \mathrm dt' C_{xx'}(t-t')  
U_S(t,t')  K_{x'\sigma},
\hspace{0.5cm}
\widehat \Lambda_{xx'\sigma}(t) = \int_{t_0}^t \mathrm dt' C_{x'x}^*(t-t') 
 U_S(t,t')  K_{x'\sigma}~.
\end{equation}
Here we employed the definitions $U_S(t,t')=T_+\exp\{-i \int_{t'}^t \mathrm
d\tau \mathcal L_S(\tau)\}$ and assumed the reservoir correlation functions
$C_{x x'}(t)=\mathrm{tr}_R \lbrace {\rm e}^{i H_R t} \Phi_x {\rm e}^{-i H_R
  t} \Phi_{x'} \rho_R \rbrace$ with the reservoir density matrix $\rho_R$
to be spin-independent.  Using this form has the advantages over the
Redfield approach \cite{redf57} that the memory terms are included and that
the wire-laser coupling is treated exactly within the dipole approximation.

In the system-bath approach used the coupling to the fermionic reservoirs
is described by a single quantity, namely the spectral density $
J_{R}(\omega)=\sum_q \pi \vert V_q \vert^2 \delta(\omega-\omega_q)$.  In
principle this coupling can be different for the right and the left lead
but we assume it to be equal for both cases. For a dense spectrum of
reservoirs the sum becomes a smooth function. To be able to use the theorem
of residues to determine the bath correlation functions, one 
approximates
the spectral density  by a numerical decomposition into
Lorentzian functions \cite{meie99,wela05a}
\begin{equation} \label{equ:spectralnum}
J_{R}(\omega)=\sum_{k=1}^m \frac{p_k}{4 \Omega_k}  
\frac{1}{(\omega -\Omega_k)^2+\Gamma_k^2},
\end{equation}
with real fitting parameters $p_k$, $\Omega_k$ and $\Gamma_k$.  Furthermore
one uses the fact that $J_R(\omega)\approx 0$ for $\omega \le 0$ which is
fulfilled for a wide range of parameters $\Omega_k$ and $\Gamma_k$.
With the  roots of $n_F$ and (\ref{equ:spectralnum}), the
application of the theorem of residues results in
\begin{eqnarray} \label{bath12dec}
C_{12}(t)&=&\sum_{k=1}^m \frac{p_k}{4 \Omega_k \Gamma_k}
\left(n_F(-\Omega_k^- +E_F) e^{-i\Omega_k^- t} \right) 
-\frac{2i}{\beta} \sum_k^{m'} J_{R}(\nu_k^\ast) e^{-i \nu_k^\ast t}
=\sum_{k=1}^{m+m'} a_{12}^k e^{\gamma_{12}^k t} \\
C_{21}(t)&=& \sum_{k=1}^m \frac{p_k}{4 \Omega_k \Gamma_k}
\left(n_F(\Omega_k^+-E_F)  e^{i\Omega_k^+ t} \right) 
-\frac{2i}{\beta} \sum_k^{m'} J_{R}(\nu_k) e^{i \nu_k t}
=\sum_{k=1}^{m+m'} a_{21}^k e^{\gamma_{21}^k t}
\end{eqnarray}
with the abbreviations $\Omega_k^+=\Omega_k+i \Gamma_k$ and
$\Omega_k^-=\Omega_k-i \Gamma_k$ and the Matsubara frequencies
$\nu_k=i\frac{2\pi k + \pi}{\beta} +E_F$.  Rigorously, the sum over the
Matsubara frequencies would be infinite but it can be truncated at a finite
$m'$ depending on the temperature of the system $T$ and the spectral width
of $J_R(\omega)$.  The pure exponential dependence of the correlation
function on time allows one to derive a set of differential equations for
the auxiliary density operators
\begin{eqnarray}\label{equ:diffaux}
\frac{\partial}{\partial t} \Lambda_{xx'\sigma}^k(t)&=& a_{xx'}^k  K_{x'\sigma}
\rho_S(t) -i [H_S(t), \Lambda_{xx'\sigma}^k(t)] + \gamma_{xx'}^k
\Lambda_{xx'\sigma}^k(t), \\
\frac{\partial}{\partial t}{\widehat\Lambda}_{xx'\sigma}^k(t)&=&\left(a_{x'x}^k\right)^\ast \rho_S(t) K_{x'\sigma}   -i [H_S(t), \widehat \Lambda_{xx'\sigma}^k(t)] + \left(\gamma_{x'x}^k \right)^\ast \widehat \Lambda_{xx'\sigma}^k(t)
\end{eqnarray}
with ${\Lambda}_{xx'\sigma}(t)=\sum_{k=1}^{m+m'} {\Lambda}_{xx'\sigma}^k(t)$ and
${\widehat\Lambda}_{xx'\sigma}(t)=\sum_{k=1}^{m+m'}
{\widehat\Lambda}_{xx'\sigma}^k(t)$.

Using the electron number operator of the left lead with the summation
performed over the reservoir degrees of freedom $N_l=\sum_{q\sigma}
c_{q\sigma}^{\dagger} c_{q\sigma}$ one can now derive an expression for the
current \cite{wela05a}
\begin{equation}\label{equ:finalcurrent}
I_l(t)=e\frac{\mathrm d}{\mathrm dt} \mathrm{tr} \, \lbrace N_l \rho_S(t)
\rbrace =-ie \, \mathrm{tr} \left\{ [N_l,H(t)] \rho_S(t) \right\}
=2e \, \mathrm{Re} \left(\mathrm{tr}_{S} \left\{ c_{1\sigma}^\dagger \Lambda_{12\sigma}(t) -c_{1\sigma}^\dagger \widehat\Lambda_{12\sigma}(t) \right\} \right)~.
\end{equation}

\section{Results}

In this section results are shown for a wire consisting either of one or
two sites. The parameters for both cases are the same.  The energy scale
used is the tight-binding hopping parameter $\Delta$ which is set to 0.1
eV. For the case of one site this is of course an artificial parameter
since it does not play a role in this particular system. Nevertheless we
use this energy scale since it is used in other work as well.  The
temperature is set to $T=0.25 \Delta \approx 290$ K.  For simplicity a
simple spectral density is used which consists only of one Lorentzian in
Eq.~(\ref{equ:spectralnum}). The parameters are chosen to be $p_1$=50 eV,
$\Gamma_1$=50 eV, and $\Omega_1$=5 eV which results in a maximum coupling
strength between wire and leads of $0.1$~$\Delta$. For these parameters
the condition $J_R(\omega)\approx 0$ for $\omega \le 0$ is nicely
fulfilled. The site energies are $E_1=E_2=50$ eV and therefore located at
the maximum of the coupling to the leads.  The Fermi energies are set to
$E_{F,l}=50.2$ eV $E_{F,l}=49.8$ eV leading to a bias voltage of 0.4 eV.
The laser pulse used in all calculations has a Gaussian shape $E(t)=A \,
\mathrm{exp} \left(-(t-T)^2/(2 \sigma^2) \right) \mathrm{sin}
\left(\omega_d t \right)$ with $\sigma=100$ fs, $T$= 400 fs and a maximum
amplitude of $A=2.405$ eV. This amplitude corresponds to a full CDT
\cite{lehm03a,wela05a}. The carrier frequency is set to 1~eV.  With the
chosen energy settings, a time unit in the system corresponds to 0.66 fs.
Using Eq.\ (\ref{equ:finalcurrent}) the current unit can be extracted to be
$I_0=2.43 *10^{-4}$~A .

\begin{figure}
\centerline{
\includegraphics[width=5.cm,clip]{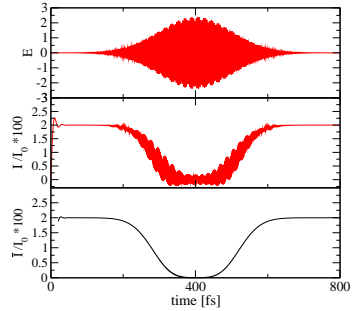}}
\caption{The scenario of CDT for a Gaussian laser field for a system with
  one site and without electron interaction. The top panel shows the laser
  field, the middle one the current through the wire, and the lowest panel
  the averaged current.}
\label{f.1}
\end{figure}

\begin{figure}
\centerline{
\includegraphics[width=5.0cm,clip,angle=270]{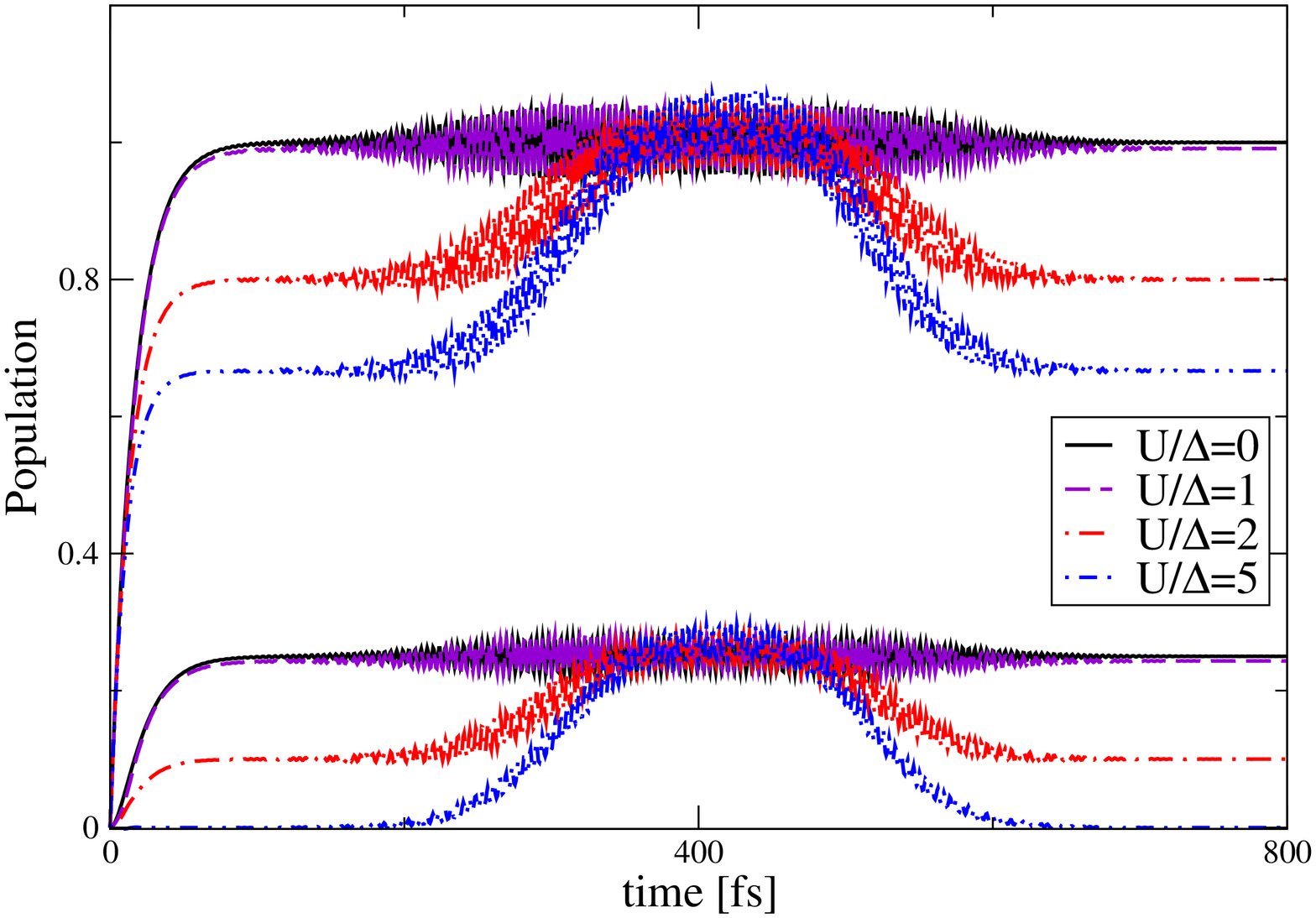}
\hspace*{1cm}
\includegraphics[width=4.9cm,clip,angle=270]{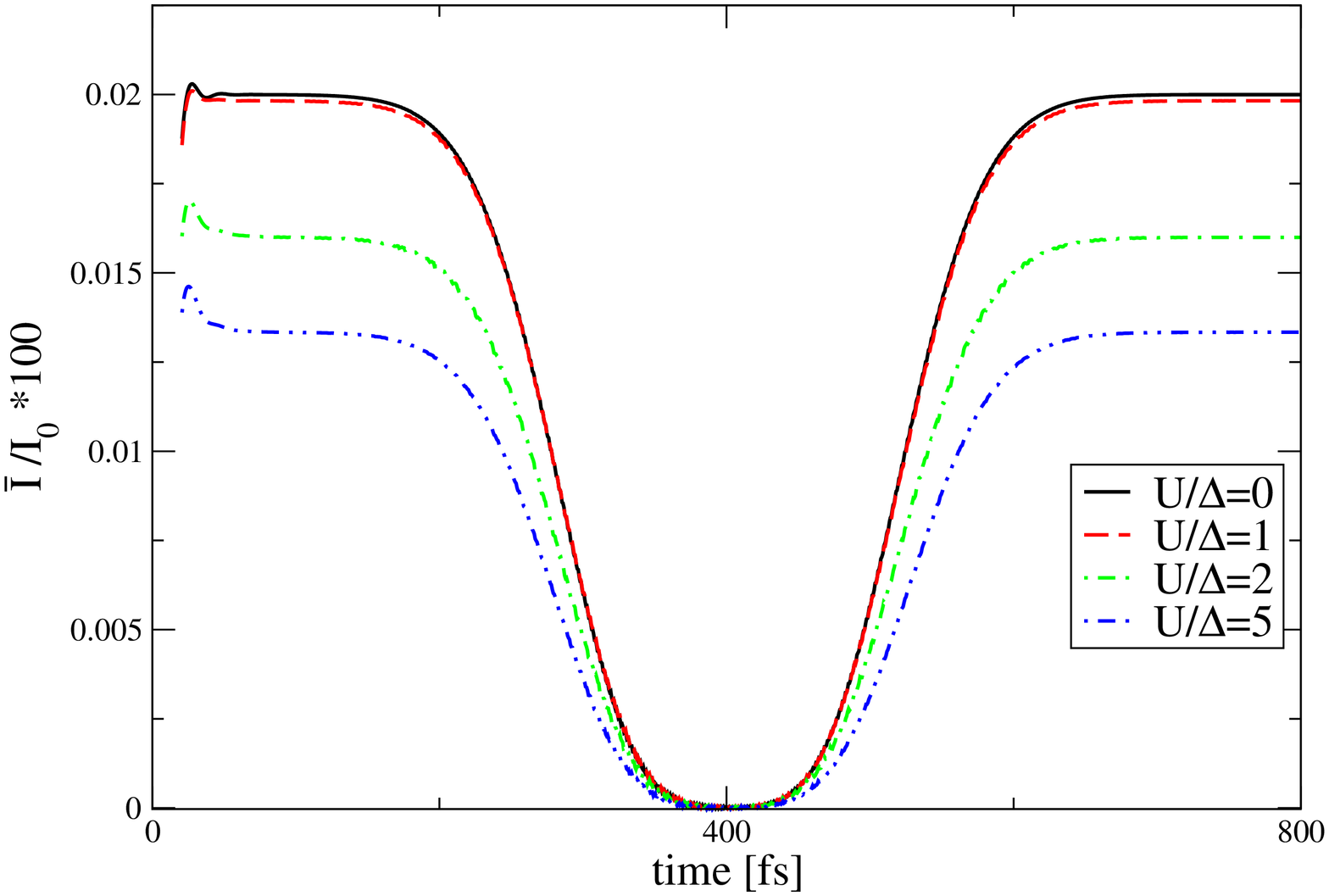}}
\caption{For a wire with only one site the population dynamics (left panel)
  and the averaged current (right panel) is shown for different values of
  electron interaction $U$. In the graph displaying the populations the
  upper four curves correspond to the population of the spin up orbital
  which is identical to the population of the spin down orbital. The lower
  four curves show the probability of a double occupancy of the site. 
}
\label{f.2}
\end{figure}

Fig.~\ref{f.2} In Fig.~\ref{f.1} the current and the averaged current are
shown together with the laser field. The amplitude $A=2.405$ eV of the
laser field is chosen such that the CDT condition is fulfilled. The CDT
condition was originally developed for cw laser fields but as already shown
earlier in the case of spinless electrons \cite{klei06b} is also important
for short laser pulses. In the case of no electron correlation as shown in
Fig.~\ref{f.1} no big difference can be seen compared to the case of
spinless electrons. For very short times one can see the loading of the
initially empty wire sites, then an equilibration before the laser pulse
sets in.  At the maximum of the laser field the average current is
completely suppressed. Interestingly Fig.~\ref{f.1} shows a wire consisting
only of one site and a constant dipole moment. So the laser suppresses the
transport between the site and the leads which are directly coupled to it.
Earlier studies always used wires of at least two sites and in those studies
it was not clear if the laser blocks the current between the sites of the
wire or between the wire and the leads. The present study clearly shows that
the CDT also works to suppress the current between the wire and the leads.

Fig.~\ref{f.2} shows the occupation probabilities and average current for
different values of the electron interaction $U$. Shown are the
probabilities of finding a spin-up electron $tr(c_{1\uparrow}^\dagger
c_{1\uparrow} \rho_S)$ which is equal to that of finding a spin-down
electron. In addition the probability of a doubly occupied site displayed.
While increasing the electron interaction $U/\Delta$ from 0 to 5 the
probability for a doubly occupied site decreased as expected. The
probability to find a spin-up or -down electron on the site decreases at
the same time since this contains also the probability for double
occupancy. Interestingly enough, the laser pulse  allows for a doubly
occupied site to an amount which is also seen without electron correlation.
So during the maximum of the laser pulse the double occupancy is the same
irrespective of electron interaction. Therefor the energy barrier created by
electron interaction is, in a sense, overcome by the energy of the laser
pulse. So it is not astonishingly that the current through the wire is the
same during the maximum of the laser pulse. But since at that moment in
time the CDT condition is fulfilled, the current is zero independent of
electron correlation. The equilibrium current without laser pulse is
smaller in the case of electron interaction since the energy of the doubly
occupied state is moved above the Fermi level due to correlation. So this
level can not participate in the current when the laser is off.

\begin{figure}
\centerline{
\includegraphics[width=5.0cm,clip,angle=270]{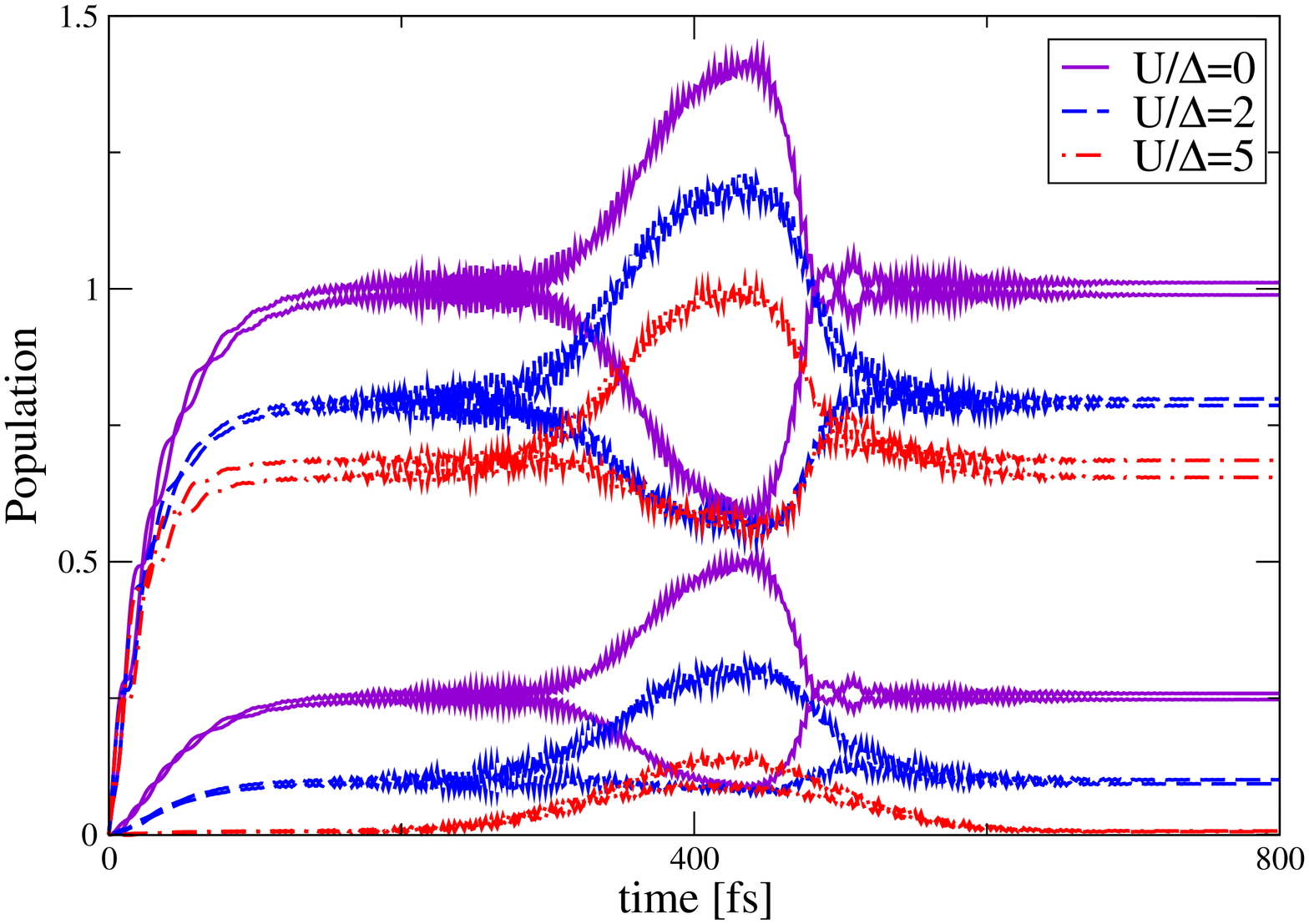}
\hspace*{1cm}
\includegraphics[width=4.9cm,clip,angle=270]{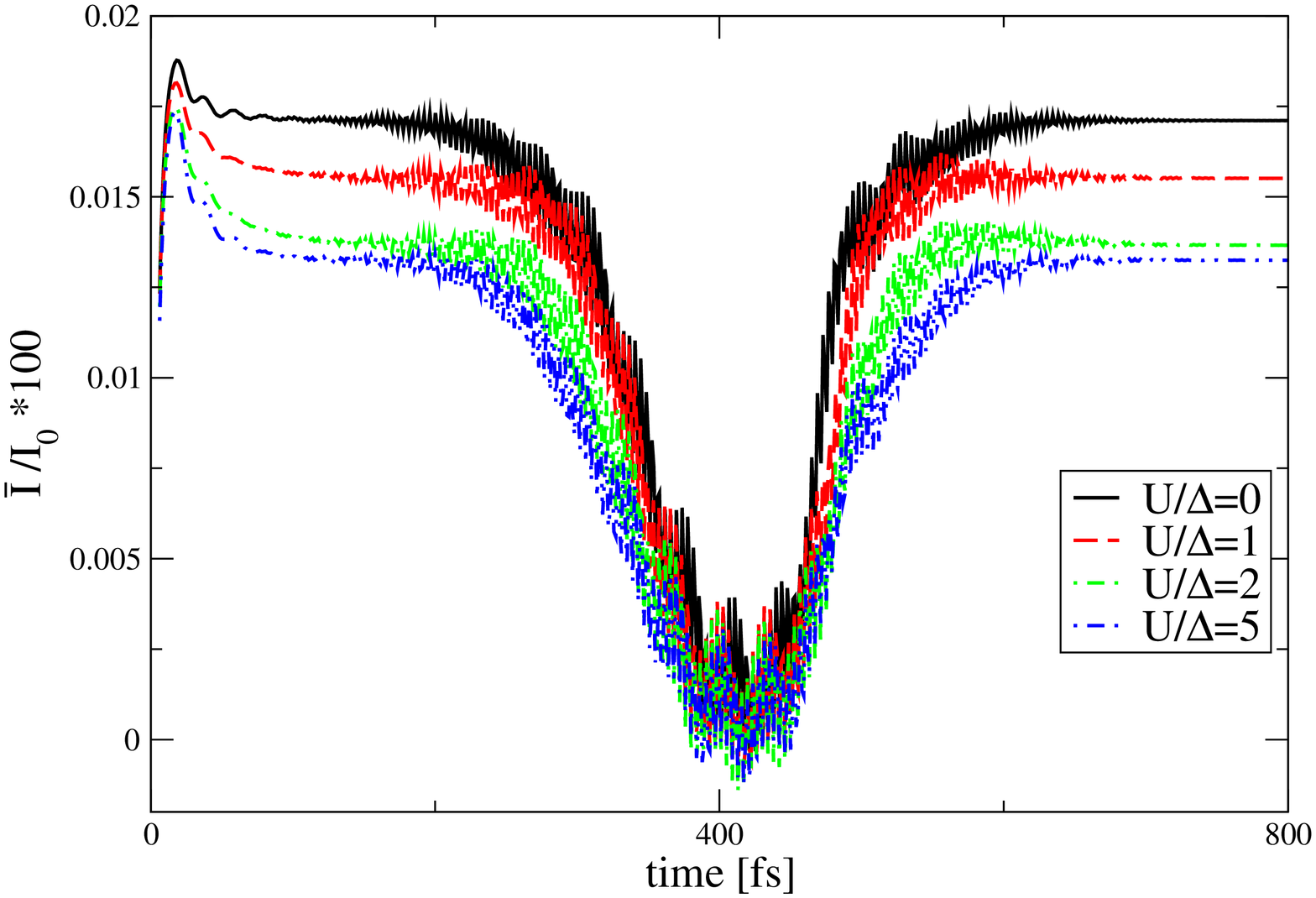}}
\caption{The population and average current through a molecular wire
  consisting of two sites. Other parameters are the same as in Fig.~3.}
\label{f.3}
\end{figure}

The case of a wire with two sites is shown in Fig.~\ref{f.3}. Again the
occupation probabilities of the different orbitals are shown. This time the
dipole operator is given by Eq.~(\ref{eq:1}) and so the population on the
two sites is different. Also in this case the probability of doubly
occupied sites decreases with electron interaction but can be increased or
decreased by the laser pulse which only shifts the site energies. The
current suppression through the phenomenon of CDT looks very similar to that
of a one-site wire or that of spinless electrons  \cite{klei06b}.


\section{Conclusions}
In the present contribution we extended our earlier work \cite{klei06b} by
including spin to be able to treat on-site interaction and by investigating
the case of a wire with only one site. Again it could be shown that the
phenomenon of CDT exists in models of molecular wires also for short laser
pulses and not only for monochromatic laser fields. The amplitude condition
as for cw-laser fields plays an important role but has to be investigated
further. As shown earlier \cite{klei06b} longer pulses  suppress the
current more effectively so it might be possible to find a better modified CDT
condition for short Gaussian laser pulses or to find  more effective pulse
forms. The on-site interaction between the electrons does not effect the
effect of current suppression which can be understood, at least partially,
by looking at the occupation probabilities and the effect of the laser
pulse on them.

The employed model of a molecular wire is of course rather simplistic.
Higher excited states should be taken into account and also dephasing
effects. The dephasing effects can be studied by coupling the wire to an
additional thermal bath in the same spirit than the coupling to the
electron reservoirs. This will be done in the future to investigate how
quickly the coherent effect of CDT gets washed out. For cw-laser fields
that has been studied within the Floquet approach \cite{lehm04a}. It is
unclear in how far short laser pulses might work better than cw-laser
fields because they do contain a range of amplitudes and so some of the
amplitudes might still fulfill a modified CDT criterion under the influence
of a thermal bath.

\end{document}